\documentclass[]{bytedance_seed}

\usepackage{amsmath}
\usepackage{amssymb}
\usepackage{amsfonts}
\usepackage{siunitx}

\title{Bridging Quantum Mechanics to Liquid Properties via a Universal Organic Force Field}

\author[1, \dagger]{Tianze Zheng}
\author[1]{Xingyuan Xu}
\author[2]{Zhi Wang}
\author[2]{Zhenze Yang}
\author[1]{Yuanheng Wang}
\author[1]{Xu Han}
\author[1]{Lei Chen}
\author[1]{Zhenliang Mu}
\author[1]{Ziqing Zhang}
\author[1]{Siyuan Liu}
\author[2]{Sheng Gong}
\author[1]{Kuang Yu}
\author[2, \dagger]{Wen Yan}

\affiliation[1]{ByteDance Seed, Beijing, China}
\affiliation[2]{ByteDance Seed, Bellevue, WA, USA}
\contribution[\dagger]{Corresponding authors}

\abstract{Molecular dynamics simulations are essential tools for unraveling atomic-level insights into the structure and behavior of condensed-phase systems. However, the universal and accurate prediction of macroscopic properties based on quantum mechanical calculations remains a significant challenge, often hindered by the trade-off between computational cost and simulation accuracy. Here we present ByteFF-Pol, a polarizable force field parameterized by a graph neural network and trained exclusively on high-level quantum mechanical data. By leveraging physically-motivated force field forms and training strategies, ByteFF-Pol predicts thermodynamic and transport properties for a wide range of small-molecule liquids and electrolytes with high accuracy, surpassing current classical and machine learning force fields. This ability to make predictions without system-specific training bridges the gap between microscopic calculations and macroscopic liquid properties, enabling the exploration of previously intractable chemical spaces. This advancement enables the precise design of new electrolytes and custom-tailored solvents, establishing a robust foundation for data-driven materials discovery.}

\date{2026.8.27}
\correspondence{Tianze Zheng at \email{zhengtianze@bytedance.com}, Wen Yan at \email{wen.yan@bytedance.com}}

\begin{document}
\maketitle

\section{Introduction}
Molecular dynamics (MD) simulations have emerged as one of the most important cornerstones of modern materials and biological research, offering an indispensable tool for probing condensed-phase systems at the atomic level~\cite{kholmurodov2007molecular,hollingsworth2018molecular,yan2023molecular}.
By revealing microscopic details of complex phenomena such as structural rearrangements and molecular diffusion—often at resolutions inaccessible to experiments—MD simulations provide critical insights for applications ranging from drug discovery to material engineering~\cite{reymond2012exploring,tafrishi2022molecular,alonso2006combining}.
Among these applications, MD simulations play a vital role in the screening and design of organic molecules, such as electrolytes for novel batteries~\cite{golovizninaExtensionCLPolPolarizable2021,gong2025predictive} and deep eutectic solvents for green chemistry~\cite{https://doi.org/10.1002/wcms.1598}.
They enable the rapid exploration of vast chemical and formula spaces for organic materials, which is impossible to enumerate through experiments.

The accuracy of these simulations, however, is critically dependent on the force field—a mathematical model describing interatomic interactions.
This dependency necessitates the development of universal force fields combining accuracy and efficiency.
Traditional force fields, such as Amber~\cite{hornak2006comparison}, CHARMM~\cite{brooks2009charmm}, and OPLS~\cite{jorgensen1984optimized,jorgensen1986optimized}, model interatomic interactions using a set of simple and predefined functional forms and tabulated parameters.
Conventional wisdom indicates that the accuracy of traditional force fields for macroscopic properties essentially relies on error cancellation at microscopic level, a consequence of their limited expressive power.
To achieve this cancellation, the force field parameters derived from low-level quantum mechanics (QM) calculations need to be further optimized by experimental data, including spectroscopic measurements~\cite{han2025refining}, and thermodynamic properties such as densities and evaporation enthalpies~\cite{jorgensen1984optimized,jorgensen1986optimized}. 

In contrast, machine learning (ML)-based force fields utilize the powerful fitting capabilities of neural networks to directly learn interatomic interactions from QM calculations, thereby achieving high accuracy in predicting intramolecular energy landscapes, spectroscopic properties, and related phenomena~\cite{kovacsEvaluationMACEForce2023,kovacsMaceOff2025,juSevenNet2025,pleFoundationModelAccurate2025}.
However, comparing to traditional force fields, ML force fields without physical constraints require tremendous QM data for training.
The development of a universal many-body ML force field is exceptionally challenging due to the vast chemical space of organic molecules and the daunting computational cost of generating sufficient ab initio data.
Consequently, prominent ML force fields like MACE-OFF can suffer from accuracy and transferability issues, resulting in inferior performance in bulk property predictions when compared to well-established traditional ones~\cite{kovacsMaceOff2025}.
While other efforts, such as the BAMBOO force field~
\cite{gong2025predictive}, have demonstrated state-of-the-art (SOTA) performance in predicting the bulk properties of electrolytes, they still depend on experimental density data for empirical fine-tuning~\cite{gong2025predictive}.
Ultimately, the predictive capability of both traditional and ML force fields remains constrained by the availability and integration of experimental information.

To address this challenge, efforts have been made toward establishing ab initio force fields by refining traditional functional forms to better incorporate the underlying physics of molecular interactions~\cite{xu2018perspective,hortonQUBEKitAutomatingDerivation2019}.
These force fields are typically based on separating the intermolecular interactions into distinct components~\cite{schmidt_transferable_2015,liu2019amoeba+} or employing many-body expansion~\cite{beran2009approximating}.
Despite the solid physical foundation, such force fields often struggle with computational efficiency and transferability~\cite{xu2018perspective}.
Polarizable force fields, including AMOEBA~\cite{ponderCurrentStatusAMOEBA2010,chungClassicalExchangePolarization2022,liu2019amoeba+}, APPLE\&P~\cite{borodinPolarizableForceField2009} and CL\&Pol~\cite{golovizninaExtensionCLPolPolarizable2021}, have emerged as a successful framework that retains computational efficiency while capturing the response of electron density to different electrical environments—a critical feature for systems like electrolytes~\cite{kirshCriticalEvaluationPolarizable2024}. 
Efforts have been made to automate the parameterization of these polarizable force fields to avoid labor-intensive process~\cite{poltype2,D5DD00178A}.
However, these techniques often handle different chemical species independently, and additional manual adjustments are sometimes required.
As a result, their transferability across different chemical environments is uncertain, failing to consistently outperform traditional force fields when applied to systems with varying electronic structures or bonding characteristics. 

In summary, currently there is still an urgent need for a universal organic force field that possesses both the accuracy of high-level ab initio calculations, as well as running speed and generalizability required by large-scale simulations of complex materials.
In response to this challenge, we introduce ByteFF-Pol, a graph neural network~\cite{corso2024graph} (GNN)-parameterized polarizable force field that represents a significant advancement in force field development.
Unlike traditional and most ML-based force fields, ByteFF-Pol is trained exclusively on high-level QM labels, eliminating the need for experimental calibration.
Through a novel force field architecture and training scheme, ByteFF-Pol achieves good accuracy in predicting the thermodynamic and transport properties of small-molecule liquids and electrolytes, outperforming SOTA traditional and ML force fields.
Its zero-shot prediction capabilities enable the exploration of previously intractable chemical spaces, opening up new avenues for data-driven discovery of materials with optimized properties for applications in electrolyte design, complex solvents, and beyond.

\section{Results}
\subsection{Overview of ByteFF-Pol}
\label{sec:overview}

\begin{figure*}[ht]
    \centering
    \includegraphics[width=\linewidth]{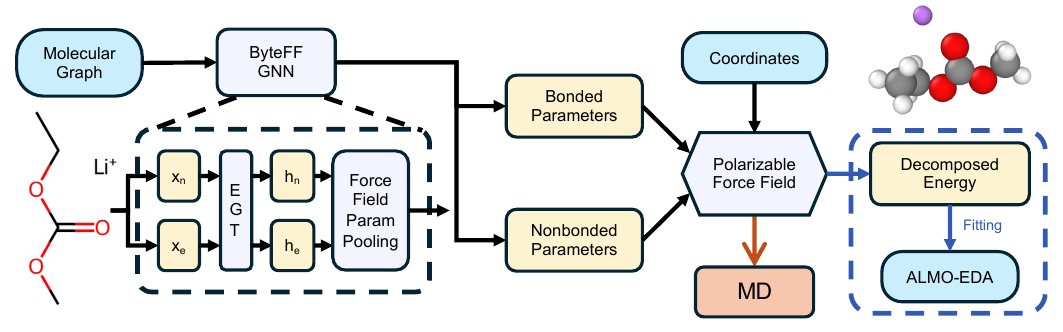}
    \caption{\textbf{Framework of ByteFF-Pol.} Starting from molecular graphs, the ByteFF graph neural network (GNN) model extracts initial atom ($x_\mathrm{n}$) and bond ($x_\mathrm{e}$) embeddings. These are processed through a multi-layer edge-augmented graph transformer (EGT) to generate refined representations ($h_\mathrm{n}$ and $h_\mathrm{e}$), which are subsequently mapped to bonded and non-bonded force field parameters. During training, these parameters are combined with atomic coordinates to compute decomposed energy terms, which are fitted against reference labels derived from the absolutely localized molecular orbital energy decomposition analysis (ALMO-EDA) calculations. In the inference stage, the model predicts force field parameters in a single forward pass for use in molecular dynamics (MD) simulations.
    }
    \label{fig:framework}
\end{figure*}

The ByteFF-Pol framework, depicted in Figure \ref{fig:framework}, utilizes a GNN model to derive force field parameters directly from molecular graphs.
This modern parameterization technique offers significant advantages over traditional look-up tables~\cite{wangEndtoendDifferentiableConstruction2022,thurlemann2023regularized,zheng2025data,seute2025grappa}.
During training, the predicted force field parameters are integrated with atomic coordinates into energy functions to compute the decomposed energy terms.
And the GNN model is optimized by fitting the decomposed energy terms to reference data from the absolutely localized molecular orbital energy decomposition analysis (ALMO-EDA) calculations~\cite{khaliulinEDAv1Original2007}.
Once trained, the finalized GNN model can generate parameters for direct use in standard MD engines (e.g., OpenMM~\cite{eastman2017openmm}) to execute MD simulations.
This approach streamlines the parameterization workflow and enhances the transferability of the force field across diverse chemical environments.

Similar to ByteFF~\cite{zheng2025data}, the GNN model is composed of three layers. 
First, a feature layer extracts information about atoms and bonds from molecular graphs to construct atom and bond embeddings ($x_\mathrm{n}$ and $x_\mathrm{e}$).
Next, these embeddings are propagated through a multi-layer edge-augmented graph transformer (EGT)\cite{hussainGlobalSelfAttentionReplacement2022} to produce the hidden representations of atoms and bonds ($h_\mathrm{n}$ and $h_\mathrm{e}$), describing local chemical environments. 
Finally, a pooling layer processes these hidden representations to generate bonded and non-bonded force field parameters. 
The GNN model carefully considers molecular symmetries in its 2D topology, so the predicted force field parameters maintain these symmetries.

The ByteFF-Pol force field energy function, $U^\mathrm{FF}$, is partitioned into bonded and non-bonded components:
\begin{equation}
    U^\mathrm{FF} = U^\mathrm{FF}_{\mathrm{bonded}} + U^\mathrm{FF}_{\mathrm{non-bonded}}
\end{equation}
The bonded energy term, $U^\mathrm{FF}_\mathrm{bonded}$, is consistent with the functional forms used in ByteFF and GAFF2, encompassing standard bond, angle, proper dihedral, and improper dihedral potentials. The non-bonded energy, $U^\mathrm{FF}_\mathrm{non-bonded}$, consists of five components: repulsion ($U^\mathrm{FF}_{\mathrm{rep}}$), dispersion ($U^\mathrm{FF}_{\mathrm{disp}}$), permanent electrostatics ($U^\mathrm{FF}_{\mathrm{est}}$), polarization ($U^\mathrm{FF}_{\mathrm{pol}}$), and charge transfer ($U^\mathrm{FF}_{\mathrm{ct}}$) terms:
\begin{equation}
    \begin{split}
    U^\mathrm{FF}_{\mathrm{non-bonded}} = & U^\mathrm{FF}_{\mathrm{rep}}(\bm{r} ; \epsilon^\mathrm{rep}, \lambda^\mathrm{rep}, r^*) + U^\mathrm{FF}_{\mathrm{disp}}(\bm{r} ; C_6,  r^*) + U^\mathrm{FF}_{\mathrm{est}}(\bm{r} ; q) \\
    &+ U^\mathrm{FF}_{\mathrm{pol}}(\bm{r} ; q, \alpha) + U^\mathrm{FF}_{\mathrm{ct}}(\bm{r} ; \epsilon^\mathrm{ct}, \lambda^\mathrm{ct}, r^*).
    \end{split}
\end{equation}
Here, $\bm{r}$ is the atomic coordinates and $r^*$, $\epsilon^\mathrm{rep}$, $\lambda^\mathrm{rep}$, $C_6$, $q$, $\alpha$, $\epsilon^\mathrm{ct}$ and $\lambda^\mathrm{ct}$ are force field parameters predicted by the GNN model.
The specific formulations of these terms are provided in the Methods section.
Crucially, this non-bonded decomposition is designed to align with the energy components given by the ALMO-EDA method.
This alignment allows the force field parameters to be trained by fitting the decomposed energy terms to the corresponding references derived from the ALMO-EDA calculations. 

ByteFF-Pol shares a similar GNN model with our previous work, ByteFF\cite{zheng2025data}, for predicting force field parameters.
The major advancements lie in the parametrization methodology of non-bonded parameters: 
instead of distilling the conventional GAFF2 force field, we now fit to QM data directly.
This constitutes a non-trivial task, as most SOTA force fields require experimental fine-tuning due to their inaccurate function form and incomplete QM dataset.
To address these challenges, we curated a large-scale, chemically diverse QM dataset, generating decomposed energy labels using the ALMO-EDA method in addition to the total energy.
The EDA data encodes physically meaningful information that substantially enhances the robustness of the resulting model.
Correspondingly, we designed EDA-compatible force field functions and a tailored training protocol to ensure both fitting efficiency and accuracy. 
In the subsequent sections, we detail the training procedure of ByteFF-Pol and demonstrate its performance in both microscopic energy predictions and bulk simulations through comprehensive benchmarking studies.

\subsection{Fitting to ALMO-EDA labels}
\label{sec:qm}

To generate accurate training labels, density functional theory (DFT) calculations at the $\omega$B97M-V/def2-TZVPD~\cite{mardirossian2016omegab97m} level are employed.
This functional is a range-separated hybrid meta-GGA that incorporates VV10 nonlocal correlation, which has been validated to achieve chemical accuracy for non-covalent interactions relative to CCSD(T) benchmarks, ensuring high fidelity for the dispersion-dominated interactions targeted in this study~\cite{mardirossian2016omegab97m,liang2025gold}.
Interaction energies between molecule dimers are further split into physically interpretable components using energy decomposition analysis (EDA).
Popular EDA approaches include Symmetry-Adapted Perturbation Theory (SAPT)~\cite{szalewicz2012symmetry}, Natural Energy Decomposition Analysis (NEDA)~\cite{glendening1996natural}, ALMO-EDA~\cite{khaliulinEDAv1Original2007}, etc.
In this study, we utilize the second-generation ALMO-EDA method to generate training labels, chosen for its clear physical interpretation and compatibility with standard DFT frameworks.

\begin{figure*}[ht]
    \centering
    \includegraphics[width=\linewidth]{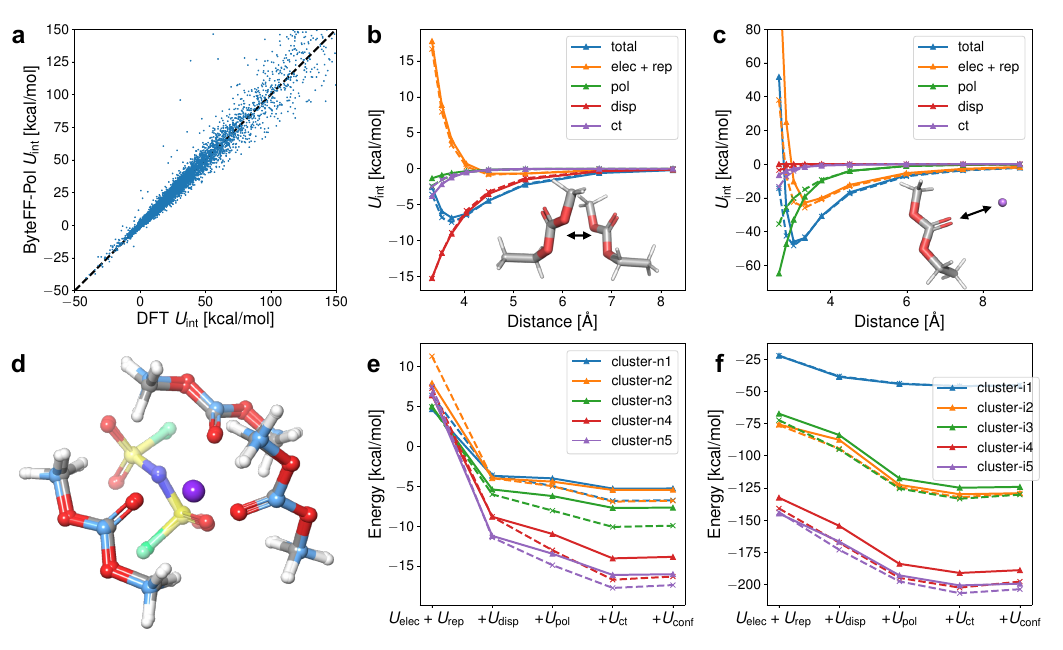}
    \caption{
        \textbf{ByteFF-Pol is validated against DFT references.} (a) Comparison of interaction energies predicted by ByteFF-Pol versus DFT on the validation set (a random sampled subset of 76 k conformations). (b, c) Dimer scan of (b) two EMC molecules and (c) an EMC molecule with a lithium ion (solid lines for ByteFF-Pol and dashed lines for DFT). (d) Cluster of a lithium ion surrounded by three DMC molecules and a FSI$^-$ anion. The conformations were first relaxed by DFT (light blue carbon) and then by ByteFF-Pol (grey carbon). (e, f) Decomposed binding energies of clusters with neutral molecules only (e) and clusters with neutral molecules and ions (f). The energy terms are accumulated so that the final point equals to the overall binding energy (solid lines for ByteFF-Pol and dashed lines for DFT).
    The corresponding components and conformations of the clusters are provided in Supplementary Note 2.1. The full names and SMILES strings corresponding to the molecular abbreviations are listed in Supplementary Table 1.
    }
    \label{fig:qm}
\end{figure*}

During training, ByteFF-Pol predicts the decomposed interaction energies of molecular dimers on the training set.
These predictions are then fitted to the corresponding EDA labels to optimize the GNN model parameters.
After training, the performance of ByteFF-Pol is verified on the validation set (Figure\ref{fig:qm} (a)).
Despite the fixed functional form, ByteFF-Pol's predictions show reasonable agreement with the DFT labels, yielding a mean absolute error (MAE) of 0.36 kcal/mol for the total interaction energy (excluding unphysical configurations above 100 kcal/mol).
The discrepancies in the lower interaction energy regions are minor, whereas those in higher interaction energy (i.e., highly repulsive) regions are comparatively large.
This error distribution is an intentional consequence of the scaled mean squared error (SMSE) loss function (Eq.~\ref{eq:SMSE}).
By incorporating Boltzmann weighting, the training prioritizes accuracy in the low-energy regions most relevant to condensed-phase equilibrium properties.

To illustrate the fitting performance for the decomposed energy terms, two dimer scan examples are presented in Figure~\ref{fig:qm}(b) and (c).
The first example shows the interaction energy between two neutral molecules as a function of separation distance.
In this system, the interactions are relatively weak and dominated by dispersion.
Using a damped $C_6$ form, ByteFF-Pol reproduces the dispersion interaction.
The second example illustrates the interaction between a neutral molecule and a lithium ion, where electrostatics and polarization effects play central roles.
ByteFF-Pol exhibits a slight overestimation of polarization interactions at short ranges; however, due to a cumulative cancellation of errors among the different terms, it still results in consistent representation of the total interaction energy.
For a broader statistical assessment, we evaluated the prediction residuals and group-wise performance across the validation set (Supplementary Note 7.4).
ByteFF-Pol demonstrates consistent accuracy across diverse chemical categories, yielding MAE values of 0.29--0.40 kcal/mol for neutral-neutral interactions and 0.53--1.39 kcal/mol for ion-molecule interactions, demonstrating overall consistency with DFT references.

To estimate the transferability of ByteFF-Pol from dimers to clusters, we perform geometric optimization on several representative clusters and compute binding energy, $U_\mathrm{bind}$ (Eq. S1).
The relaxed geometries are shown in Figure~\ref{fig:qm}(d) and Supplementary Note 2.1.
The resulting structures from DFT and ByteFF-Pol align well with each other, demonstrating the capability of ByteFF-Pol to predict equilibrium geometries reliable for these clusters.
The binding energy of various clusters is shown in Figure~\ref{fig:qm} (e) for neutral systems and Figure~\ref{fig:qm} (f) for clusters containing ions.
In these plots, the individual energy components are cumulatively summed so that the final value represents the total binding energy.
The neutral examples were chosen to encompass diverse interaction types, including hydrogen bonding, polar interactions, $\pi-\pi$ stacking, etc.
Similar to the dimer scan results, these interactions are relatively weak and dispersion-dominated.
ByteFF-Pol exhibits a consistent trend with DFT calculations but tends to underestimate polarization and charge transfer energies, possibly due to the complex anisotropic nature of intermolecular interactions.

In clusters containing both neutral molecules and ions, the interactions are significantly stronger. 
Nevertheless, ByteFF-Pol yields results consistent with DFT references for these clusters, which represent the local chemical environments typical of common electrolytes.
Notably, polarization interactions are inherently many-body effects, arising from the non-pairwise additive nature of electrical responses. 
Leveraging a physically-motivated self-consistent model for polarization, ByteFF-Pol maintains reliable accuracy in predicting this component even in extended clusters.
The accuracy of ByteFF-Pol across cluster potential energy surfaces (PES) is further validated by the interaction energies of configurations sampled from MD simulations (Supplementary Figure 11). We obtained MAE values of 0.11--0.41 kcal/mol for neutral clusters and 0.85--4.82 kcal/mol for ion-containing systems, with Pearson correlation coefficients near 1.0 for all clusters. These results confirm the robust transferability of ByteFF-Pol from dimer-only training to representative many-body environments.
Overall, the predictions from ByteFF-Pol for interaction energies, spanning from simple dimers to complex multi-molecule clusters, show strong concordance with the DFT benchmarks.

\subsection{Thermodynamic Properties of Molecular Liquids}
\label{sec:mol_liq}
The thermodynamic properties of molecular liquids, primarily governed by intermolecular interactions, serve as critical benchmarks for evaluating force field validity.~\cite{caleman2012force,fischer2015properties}
Using ByteFF-Pol, we computed the densities and evaporation enthalpies for a diverse set of nearly one hundred pure molecular liquids, covering a wide range of functional groups to represent different interaction types.
The performance of ByteFF-Pol is benchmarked against that of SOTA force fields, including a conventional force field (i.e. OPLS-AA~\cite{jorgensen1984optimized,jorgensen1986optimized}), a polarizable force field (i.e. AMOEBA~\cite{ponderCurrentStatusAMOEBA2010,walkerAutomationAMOEBAPolarizable2022}), and a ML force field (i.e., MACE-OFF23(M)~\cite{kovacsMaceOff2025}), as shown in Figure~\ref{fig:mol_liq} and Table~\ref{tab:mol_liq}.
Parmetrization and simulation details can be found in Supplementary Note 8, and comparsion with additional force fields are provided in Supplementary Note 9.1.

For liquid densities, ByteFF-Pol provides predictions comparable to established force fields like OPLS-AA and AMOEBA.
The minor disagreement with experimental values is mainly due to a slight, systematic overestimation.
This overestimation is consistent with the expected impact of neglecting the nuclear quantum effects (NQEs), which typically decreases the density of hydrogen-rich molecules by about 2\%~\cite{kurnikov2024neural}.
We address the NQEs by performing Path-Integral Molecular Dynamics (PIMD) simulations~\cite{tuckerman_statistical_2023}, with detailed in Supplementary Note 8.6.
As shown in Figure~\ref{fig:mol_liq}(a), ByteFF-Pol-NQE has effectively corrected the systematic overestimation of ByteFF-Pol, resulting in a lower MAE and mean absolute percentage error (MAPE) than other force fields.

Predicting the evaporation enthalpy ($\Delta H_{\text{vap}}$), however, is more challenging due to experimental uncertainties and the ideal gas approximation used for the gas phase simulations (see Supplementary Note 8.2).
Despite these difficulties, ByteFF-Pol demonstrates remarkable accuracy, outperforming both OPLS-AA and AMOEBA.
With the inclusion of NQEs, ByteFF-Pol-NQE yields even better results. 
This high accuracy is particularly noteworthy given that ByteFF-Pol is trained exclusively on DFT dimer data, whereas OPLS-AA and AMOEBA are parameterized using experimental data including densities and evaporation enthalpies. 

When compared to MACE-OFF, a SOTA ML force field trained purely on DFT data, ByteFF-Pol exhibits substantially lower prediction errors for both density and evaporation enthalpy.
As shown in Supplementary Figure 16, equipped with physics-driven long-range interactions, MACELES-OFF~\cite{kim2025universal} significantly outperforms MACE-OFF, although it still exhibits larger errors than ByteFF-Pol.
This discrepancy highlights the effectiveness of the physically-motivated model and training strategy employed by ByteFF-Pol.

\begin{figure*}[ht]
    \centering
    \includegraphics[width=\linewidth]{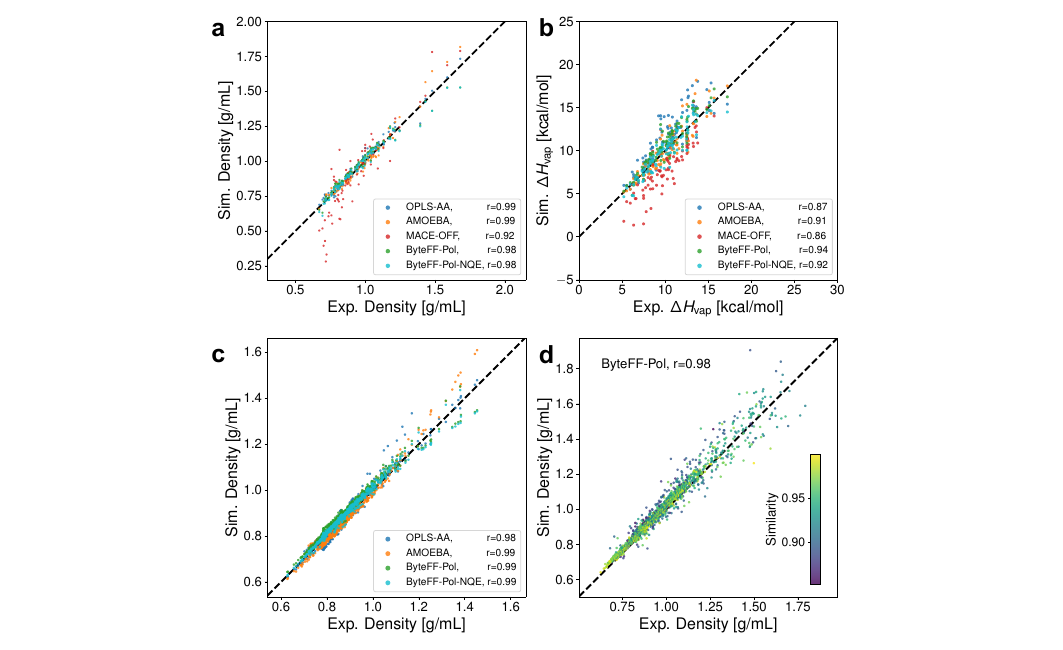}
    \caption{\textbf{Benchmarks on the thermodynamic properties of molecular liquids.} Comparison of different force fields on (a) density and (b) evaporation enthalpy ($\Delta H_{\text{vap}}$) of pure molecular liquids from QUBEKit~\cite{hortonQUBEKitAutomatingDerivation2019}; (c) density of binary molecular liquids from ThermoML~\cite{chirico2003thermoml}; and (d) density prediction by ByteFF-Pol on additional pure molecular liquids from CRC handbook~\cite{Lide2004CRC}. The results of MACE-OFF were acquired from Ref \citenum{kovacsMaceOff2025}. Pearson correlation coefficients (r) of each force field are provided in the legends.
    }
    \label{fig:mol_liq}
\end{figure*}

\begin{table}[ht]
\centering
\caption{Statistical results of thermodynamic properties of molecular liquids. The best results for each metric are highlighted in bold.}
\footnotesize
\label{tab:mol_liq}
    \begin{tabular}{c|cccc|cccc}
    \toprule
    & \multicolumn{4}{c|}{Density (g/mL) from QUBEKit~\cite{hortonQUBEKitAutomatingDerivation2019}} & \multicolumn{4}{c}{$\Delta H_\mathrm{vap}$ (kcal/mol) from QUBEKit~\cite{hortonQUBEKitAutomatingDerivation2019}} \\
    & $n_\mathrm{system}$ & MAE  & MAPE & Pearson & $n_\mathrm{system}$ & MAE  & MAPE & Pearson \\
    \midrule
    OPLS-AA & 96 & 0.024 & 2.5\% & \textbf{0.99} &  96 & 1.52 & 14.8\% & 0.87 \\
    AMOEBA & 87 & 0.025 & 2.4\% & \textbf{0.99} & 87 & 0.85 & 8.1\% & 0.91 \\
    MACE-OFF & 92 & 0.084 & 9.8\% & 0.92 & 80 & 2.14 & 23.1\% & 0.86\\
    ByteFF-Pol & 96 & 0.029 & 3.0\% & 0.98 & 96 & 0.95 & 9.3\% & \textbf{0.94} \\
    ByteFF-Pol-NQE & 96 & \textbf{0.023} & \textbf{2.3\%} & 0.98 & 96 & \textbf{0.76} & \textbf{7.5\%} & 0.92 \\
    \midrule
    & \multicolumn{4}{c|}{Density (g/mL) from ThermoML~\cite{chirico2003thermoml}} & \multicolumn{4}{c}{Density (g/mL) from CRC handbook~\cite{Lide2004CRC}} \\
    & $n_\mathrm{system}$ & MAE  & MAPE & Pearson & $n_\mathrm{system}$ & MAE  & MAPE & Pearson \\
    \midrule
    OPLS-AA & 594 & 0.018 & 2.0\% & 0.98 & - & - & - & - \\
    AMOEBA & 554 & 0.018 & 2.0\% & \textbf{0.99} & - & - & - & - \\
    ByteFF-Pol & 595 & 0.028 & 3.1\% & \textbf{0.99} & 2152 & 0.032 & 3.28\% & 0.98 \\
    ByteFF-Pol-NQE & 595 & \textbf{0.017} & \textbf{1.8\%} & \textbf{0.99} & - & - & - & - \\
    \bottomrule
    \end{tabular}
\end{table}

The accuracy of ByteFF-Pol is further validated by its density predictions for binary molecular mixtures, as shown in Figure~\ref{fig:mol_liq}(c).
In these systems, ByteFF-Pol provides reliable accuracy comparable to that of OPLS-AA and AMOEBA.
A slight, systematic overestimation of density by ByteFF-Pol can again be corrected by incorporating the NQEs.
However, due to the substantial computational cost of PIMD simulations, we did not perform PIMD simulations for the subsequent benchmarks.
Minor underestimations are observed in some high-density systems containing polyhalogenated molecules, which is likely due to the inherent limitations of the model's functional form in capturing anisotropic interactions such as halogen bonds and $\sigma$-holes.
Nevertheless, ByteFF-Pol demonstrates reliable overall performance, effectively modeling intermolecular forces across diverse chemical environments.

The transferability and universality of ByteFF-Pol are evaluated using density data for pure molecular liquids sourced from CRC handbook (Figure \ref{fig:mol_liq} (d)).
Despite being trained on fewer than 400 molecules, ByteFF-Pol delivers reasonable density predictions for over 2,000 distinct molecules. 
The data points are colored based on the similarity of the local atomic chemical environments between the tested molecules and the training set, gauged by the similarity of the embedding vectors.
As shown in Figure \ref{fig:mol_liq} (d), the densities of most molecules with large similarity to the training set can be predicted accurately by ByteFF-Pol.
As demonstrated in Supplementary Note 3.2, a clear correlation exists between the similarity score and the density prediction accuracy.
This observation underscores that ByteFF-Pol's transferability stems from its ability to capture the chemical environments of atoms, rather than explicit training on entire molecular structures. 
This indicates that comprehensive coverage of atomic chemical environments in the training set is sufficient to enable reasonable predictions across chemically diverse molecules (see Supplementary Note 10 for a detailed analysis of the training set's chemical space).

To further demonstrate the transferability and physical robustness of ByteFF-Pol, we evaluated additional thermodynamic properties including the dielectric constant ($\varepsilon(0)$), volumetric expansion coefficient ($\alpha_p$), and isothermal compressibility ($\kappa_T$).
As detailed in Supplementary Note 12.2, ByteFF-Pol yields reliable predictions across these metrics, matching or exceeding the performance of well-established classical force fields such as OPLS-AA.
The accuracy in predicting the dielectric constant, in particular, underscores the model's ability to capture complex many-body electronic responses through its explicit polarization treatment without relying on experimental parameter fitting.
Overall, ByteFF-Pol demonstrates reliable performance in predicting the thermodynamic properties of molecular liquids with promising transferability.

\subsection{Applications in Liquid Electrolytes}

One important application of organic liquid simulation is the design of electrolytes for batteries. 
In addition to thermodynamic properties, transport properties such as viscosity, ionic conductivity, and self-diffusion coefficient are also critical for electrolytes. 
For typical electrolyte systems, ByteFF-Pol is compared with two conventional force fields, OPLS-AA and OPLS-AA-SC, as well as a SOTA polarizable force field, AMOEBA. 
Instead of the original OPLS-AA charges, OPLS-AA-SC employs CHELPG charges~\cite{breneman1990determining} scaled by a factor of 0.8, an empirical charge model widely used to approximate polarization in electrolyte systems.

\begin{figure*}[ht]
    \centering
    \includegraphics[width=\linewidth]{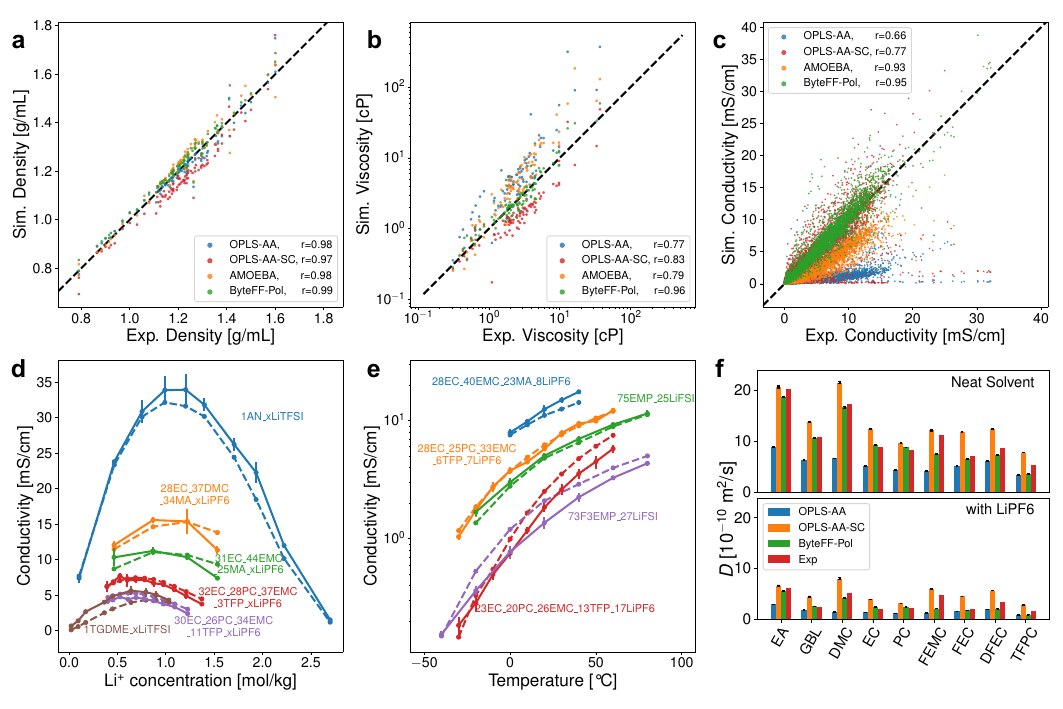}
    \caption{\textbf{Evaluation of ByteFF-Pol for predicting electrolyte properties.} (a--c) Comparison of different force fields in predicting (a) density, (b) viscosity and (c) conductivity of liquid electrolytes, respectively. Pearson correlation coefficients (r) of each force field are provided in the legends. (d, e) Predictive accuracy for conductivity as a function of (d) lithium salt concentration and (e) temperature. Solid lines represent simulation results from ByteFF-Pol, while dashed lines denote experimental reference data; numerical values in the legends indicate the approximate molar fraction of each chemical species. (f) Self-diffusion coefficient ($D$) for different species within electrolyte systems. Upper bars correspond to $D$ of species X (labeled on the x axis) in the X:EMC binary solvent (molar ratio=1:1); and lower bars show the $D$ of X in the LiPF6:X:EMC electrolytes (molar ratio=1:4:4). Experimental data are obtained from Ref~\citenum{C9EE00141G}. The full names and SMILES strings corresponding to the molecular abbreviations are provided in Supplementary Table 1. Error bars in (d--f) represent the standard deviation (s.d.) derived from four independent molecular dynamics (MD) simulations.
    }
    \label{fig:electrolyte}
\end{figure*}

\begin{table}[ht]
\centering
\caption{Statistical results of electrolyte systems. The best results for each metric are highlighted in bold.}
\label{tab:electrolyte}
    \begin{tabular}{c|cccc|cccc}
    \toprule
    & \multicolumn{4}{c|}{Density (g/mL)} & \multicolumn{4}{c}{Viscosity (cP)} \\
    & $n_\mathrm{system}$ & MAE  & MAPE & Pearson & $n_\mathrm{system}$ & MAE  & MAPE & Pearson \\
    \midrule
    OPLS-AA & 88 & 0.022 & 1.8\% & 0.98 & 91 & 12.20 & 163.3\% & 0.77 \\
    OPLS-AA-SC & 88 & 0.054 & 4.4\% & 0.97 & 91 & 1.79 & 39.8\% & 0.83 \\
    AMOEBA & 88 & 0.030 & 2.4\% & 0.98 & 90 & 6.69 & 103.4\% & 0.79 \\
    ByteFF-Pol & 88 & \textbf{0.024} & \textbf{1.9\%} & \textbf{0.99} & 91 & \textbf{1.06} & \textbf{20.0\%} & \textbf{0.96} \\
    \midrule
    & \multicolumn{4}{c|}{Conductivity (mS/cm)} & \multicolumn{4}{c}{Conductivity $>$ 2.0 mS/cm} \\
    & $n_\mathrm{system}$& MAE  & MAPE & Pearson & $n_\mathrm{system}$ & MAE  & MAPE & Pearson \\
    \midrule
    OPLS-AA & 4896 & 4.46 & 86.3\% & 0.66  & 3360 & 6.20 & 86.9\% & 0.57 \\
    OPLS-AA-SC & 4896 & 1.55 & 64.9\% & 0.77 & 3360 & 2.04 & 29.5\% & 0.63 \\
    AMOEBA & 3735 & 2.70 & 68.4\% & 0.93 & 2582 & 3.66 & 56.4\% & 0.90 \\
    ByteFF-Pol & 4896 & \textbf{0.86} & \textbf{52.4\%} & \textbf{0.95} & 3360 & \textbf{1.09} & \textbf{17.1\%} & \textbf{0.92} \\
    \bottomrule
    \end{tabular}
\end{table}

As shown in Figure~\ref{fig:electrolyte} and Table~\ref{tab:electrolyte}, OPLS-AA maintains high accuracy for density but overestimates viscosity and underestimates conductivity.
While the scaled charge approach employed by OPLS-AA-SC improves conductivity accuracy, it introduces underestimation of both density and viscosity.
These discrepancies primarily arise from the neglect of explicit polarization in conventional force fields, a factor demonstrated to be critical for electrolyte systems in the ablation study (Supplementary Note 11).
Although AMOEBA achieves better alignment with experimental reference by accounting for polarization, it still exhibits a systematic overestimation of viscosity and underestimation of conductivity.
This limitation might be attributed to its training protocol, which relies on fitting to experimental bulk data; consequently, the force field's performance may diminish when the chemical space of the electrolyte systems extends beyond the coverage of its original experimental training set.
Furthermore, AMOEBA suffers from occasional parameterization failures and simulation instabilities, resulting in unavailable conductivity data for approximately one-quarter of the tested systems.

In contrast, ByteFF-Pol successfully predicts all three properties, emphasizing the importance of explicitly modeling polarization.
On this extensive conductivity benchmark encompassing nearly 5,000 electrolyte systems with diverse components and temperature conditions, ByteFF-Pol achieves a Pearson correlation coefficient of 0.95, while yielding the lowest MAE and MAPE, demonstrating its utility as a tool for screening high-conductivity electrolytes.
For other electrolyte force fields, large-scale benchmarking of this scope was not feasible; however, comparative analyses provided in Supplementary Note 9.2 underscore ByteFF-Pol's advantages.
As shown in Supplementary Figure 17 and Supplementary Table 11, ByteFF-Pol exhibits superior accuracy in predicting conductivity compared to BAMBOO, the current SOTA machine learning force field for electrolyte systems. The experimental conductivity data presented in Supplementary Figure 17 were obtained from iterative optimization of an electrolyte system~\cite{zhu2024differentiable}. As a result of the reliable agreement between ByteFF-Pol's predictions and experimental observations, the model holds significant potential to replace such experimental measurements, thereby reducing associated costs.

To further demonstrate the utility of ByteFF-Pol in electrolyte design, we present representative examples highlighting its practical potential.
In real-world research, a key objective is to identify the optimal lithium salt concentration that maximizes electrolyte conductivity—a goal achievable by leveraging ByteFF-Pol to predict conductivity across varying lithium salt concentrations.
Examples encompass systems containing two types of lithium salts, paired with cyclic carbonates, linear carbonates, esters, ethers, and nitriles, with conductivity values spanning 0.1 to 35 mS/cm are provided in Figure \ref{fig:electrolyte} (d).
It is shown that ByteFF-Pol not only delivers reliable conductivity predictions for diverse systems but also captures the positions of conductivity peaks.

Besides, there is a growing demand for batteries capable of operating across a broad temperature range.
Accurate prediction of electrolyte conductivity under varying temperature conditions facilitates the screening of suitable electrolytes for such applications.
Representative examples of conductivity across different temperatures and diverse systems are presented in Figure \ref{fig:electrolyte} (e).
It is evident that ByteFF-Pol's predictions exhibit good agreement with experimental data over a temperature range of -40 to \SI{75}{\degreeCelsius}.
This performance highlights a key advantage of ByteFF-Pol: it is truly learned from QM potential energy surfaces rather than being empirically fitted to experimental data.

The diffusion coefficient of solvent molecules in electrolytes is a key property linked to the solvation structure of lithium ions~\cite{C9EE00141G}.
Given that the definition and derivation of solvation structures lack clear standardization, we opted to directly benchmark ByteFF-Pol's performance by comparing its predicted diffusion coefficients against experimental measurements.
Figure~\ref{fig:electrolyte}(f) presents these coefficients for various solvents in systems with and without lithium salts.
For common esters and carbonates, ByteFF-Pol's predictions exhibit reasonable agreement with experimental results.
For fluorinated solvents, ByteFF-Pol shows a systematic underestimation, which, as discussed previously, can be attributed to its limitations in capturing anisotropic interactions. Despite this, ByteFF-Pol correctly predicts the experimental trends even for these challenging systems.
In comparison, OPLS-AA consistently underestimates the diffusion coefficient across all systems, while OPLS-AA-SC consistently overestimates it.

Owing to its relatively simple functional form, ByteFF-Pol achieves a simulation speed of approximately 40 ns/day for a 10,000-atom system on an A100 GPU, and 32 ns/day on the more cost-effective A10 GPU.
This computational efficiency is markedly superior to that of typical ML force fields (see Supplementary Note 9.3 for detailed comparison).
In summary, ByteFF-Pol exhibits excellent accuracy and efficiency in predicting diverse electrolyte properties.
This capability, on one hand, validates the feasibility of developing universal force fields from ab initio calculations and, on the other hand, paves the way for exploring novel electrolytes across a vast chemical space.

\section{Discussion}
A longstanding goal in computational chemistry and materials science is to predict the macroscopic properties of condensed matter systems from ab initio computations.
However, obtaining properties such as conductivity via ab initio MD simulations remains computationally prohibitive.
Historically, this gap has been filled by traditional empirical force fields, such as OPLS-AA or AMOEBA, which rely on error cancellation and often require empirical fine-tuning to achieve macroscopic accuracy.

In this study, ByteFF-Pol serves as a critical bridge between microscopic DFT calculations and macroscopic liquid properties.
Trained exclusively on DFT dimer labels, ByteFF-Pol demonstrates robust accuracy in predicting both thermodynamic properties (e.g., density and evaporation enthalpy) and transport properties (e.g., viscosity, diffusion coefficient, and conductivity), as validated through benchmarks covering extensive chemical space.

Its performance and derivation distinguish its methodology from the current state of the art in two primary ways.
First, unlike purely local MLFFs such as MACE-OFF, which may require massive datasets or struggle with the transferability of bulk property predictions, ByteFF-Pol utilizes explicit polarization and charge-transfer terms. This physically-grounded approach allows macroscopic behavior to emerge naturally from fundamental intermolecular interactions.
Second, the GNN-parametrization methodology, combined with efficient batch training on large-scale datasets, endows it with strong transferability across the chemical space of organic molecules, warranting its reliability in predicting properties of unseen molecules.
This automated framework offers a transferable method to parameterize physical-based force fields with reliable accuracy, instead of the traditional tabulated fitting and empirical adjustments.

These advantages are pivotal to ByteFF-Pol's success, rendering it a truly universal and ab initio force field that operates without reliance on experimental data, while achieving simulation speeds (e.g., 40 ns/day for 10,000 atoms) superior to most ML-based potentials.
Its zero-shot predictive power and transferability open new avenues for research in functional liquid design, from electrolytes and deep eutectic solvents to other complex liquid systems.

While ByteFF-Pol already outperforms existing force fields, it retains potential for further optimization.
Future work will focus on enhancing its functional form to better model anisotropic interactions, thereby improving accuracy in systems containing complex local interactions.
In summary, ByteFF-Pol represents a significant step toward ab initio-guided prediction of macroscopic properties, offering a scalable, data-efficient tool for materials discovery.

\section{Methods}
\subsection{ALMO-EDA}

The total interaction energy within a multi-fragment system, $U_{\mathrm{int}}^{\mathrm{DFT}}$, is calculated at the DFT level as the difference between the self-consistent field (SCF) energy of the entire cluster ($U^{\mathrm{DFT}}[\rho_{\mathrm{full}}]$) and the sum of the SCF energy of the isolated fragments ($U_F^{\mathrm{DFT}}[\rho_F]$):
\begin{align}
U_{\mathrm{int}}^{\mathrm{DFT}} = U^{\mathrm{DFT}}[\rho_{\mathrm{full}}] - \sum_{F} U_F^{\mathrm{DFT}}[\rho_F].
\end{align}

This interaction energy is partitioned into physically meaningful components using the second-generation ALMO-EDA~\cite{khaliulinEDAv1Original2007}. 
This variational scheme decomposes the total interaction energy as follows:
\begin{align}
U_{\mathrm{int}}^{\mathrm{DFT}} = U_{\mathrm{frz}}^{\mathrm{EDA}} + U_{\mathrm{pol}}^{\mathrm{EDA}} + U_{\mathrm{ct}}^{\mathrm{EDA}}.
\label{eq:eda_total_decompose}
\end{align}
Here, $U_{\mathrm{frz}}^{\mathrm{EDA}}$ represents the frozen energy, which arises from the interaction between fragments without relaxing electronic structure.
The polarization energy, $U_{\mathrm{pol}}^{\mathrm{EDA}}$, accounts for the energy stabilization as each fragment's electron density relaxes in the electric field of its neighbors.
Finally, the charge-transfer energy, $U_{\mathrm{ct}}^{\mathrm{EDA}}$, corresponds to the energy stabilization from electron delocalization between fragments.

The frozen term is further decomposed into dispersion ($U_{\mathrm{disp}}^{\mathrm{EDA}}$), electrostatics ($U_{\mathrm{est}}^{\mathrm{EDA}}$), and Pauli ($U_{\mathrm{pauli}}^{\mathrm{EDA}}$) terms \cite{hornEDAElectrostaticDispersion2016}:
\begin{align}
U_{\mathrm{frz}}^{\mathrm{EDA}} = U_{\mathrm{disp}}^{\mathrm{EDA}} + U_{\mathrm{est}}^{\mathrm{EDA}} + U_{\mathrm{pauli}}^{\mathrm{EDA}}.
\label{eq:eda_frozen_decompose}
\end{align}

The precise mathematical definitions and implementation details of all decomposed energy terms are provided in Supplementary Note 5.

\subsection{Force Field Formulation}
\label{sec:ff}
As discussed in Section \ref{sec:overview}, the ByteFF-Pol potential energy function, $U^\mathrm{FF}$, consists of bonded terms and non-bonded terms,
\begin{align}
    U^\mathrm{FF} &= U_{\mathrm{bonded}}^\mathrm{FF} + U_{\mathrm{non-bonded}}^\mathrm{FF}.
\end{align}
The detailed bonded terms could be found in Supplementary Note 6.2.
The non-bonded energy, $U_{\mathrm{non-bonded}}^\mathrm{FF}$, comprises five terms:
\begin{align}
    U_{\mathrm{non-bonded}}^\mathrm{FF} &= U_{\mathrm{rep}}^\mathrm{FF} + U_{\mathrm{disp}}^\mathrm{FF} + U_{\mathrm{est}}^\mathrm{FF} + U_{\mathrm{pol}}^\mathrm{FF} + U_{\mathrm{ct}}^\mathrm{FF}.
\end{align}

Repulsion and dispersion interactions are described by a modified Buckingham (exp-6) potential \cite{mason_transport_1954,borodin_development_2006}, which has shown superior agreement with DFT-computed intermolecular energies compared to the Lennard-Jones potential \cite{rowley_determination_2001,borodin_development_2006}.
The functional forms are:
\begin{align}
U_{\mathrm{rep}}^\mathrm{FF}(\bm{r}; \epsilon^\mathrm{rep}, \lambda^\mathrm{rep}, r^*) &= \sum_{i<j}\left( \epsilon^\mathrm{rep}_{ij} \exp\left(\lambda^\mathrm{rep}_{ij}(1 - \frac{r_{ij}}{r^*_{ij}})\right)+\left(\frac{s_{\mathrm{rep}}}{r_{ij}}\right)^{12}\right) \\
U_{\mathrm{disp}}^\mathrm{FF}(\bm{r}; C_{6}, r^*) &= -\sum_{i<j}\frac{C_{6,ij}}{s_\mathrm{disp} {r^*_{ij}}^6+r_{ij}^6}.
\end{align}
Here, $r_{ij}$ is the interatomic distance between atoms $i$ and $j$, $\epsilon^\mathrm{rep}$, $\lambda^\mathrm{rep}$, $C_{6}$, and $r^*$ are atom-specific parameters predicted by the GNN model.
An additional $1/r^{12}$ term in the repulsion potential, scaled by a small constant $s_{\mathrm{rep}}$ (1.5 \AA $\cdot (\mathrm{kcal/mol})^{1/12}$), is included to prevent unphysical attraction at very short interatomic distances.
The dispersion term is damped at short ranges to avoid singularities and the double-counting of correlation effects, analogous to the DFT-D3 treatment \cite{grimme_consistent_2010}.
The dimensionless damping factor $s_\mathrm{disp}$ is set to 120.
As demonstrated in Section~\ref{sec:qm}, this damped dispersion function provides a reasonable fit to the target ALMO-EDA data.

The electrostatic interaction are modeled following the AMOEBA polarizable framework \cite{demerdashAssessingManybodyContributions2017}.
We use atom-centered point charges for the permanent electrostatics and an induced dipole model for polarization:
\begin{align}
U_{\mathrm{est}}^\mathrm{FF}(\bm{r}; q) &= \sum_{i<j}\frac{q_i q_j}{4\pi \epsilon_0 r_{ij}} \\
U_{\mathrm{pol}}^\mathrm{FF}(\bm{r}; q, \alpha) &= -\frac{1}{2}\sum_i \boldsymbol{\mu}_i^\mathrm{ind} \cdot \mathbf{E}^\mathrm{p}_i.
\end{align}
This simplification from a full multipole expansion to monopoles streamlines parameterization without a significant loss of accuracy.
The induced dipole, $\boldsymbol{\mu}^\mathrm{ind}$, is determined self-consistently by solving the polarization equations in response to the electric field, $\mathbf{E}^\mathrm{p}$ (Eq. S27).
For efficient training, $\boldsymbol{\mu}^\mathrm{ind}$ is computed via direct matrix inversion, as detailed in Supplementary Note 6.1.
The atomic charge $q$ and polarizability $\alpha$ used in solving $\boldsymbol{\mu}^\mathrm{ind}$ are predicted by the GNN model.

Finally, the charge-transfer term is modeled using a functional form inspired by the work of Deng et al. \cite{deng2017estimating}:
\begin{align}
U_{\mathrm{ct}}^\mathrm{FF}(\bm{r}; \epsilon^\mathrm{ct}, \lambda^\mathrm{ct}, r^*)&=-\sum_{i<j} \frac{\epsilon^\mathrm{ct}_{ij}}{r_{ij}^4} \exp \left( -\left(\frac{\lambda^\mathrm{ct}_{ij} r_{ij}}{r^*_{ij}}\right)^3 \right),
\end{align}
The GNN model predicts the charge transfer parameters $\epsilon^\mathrm{ct}$ and $\lambda^\mathrm{ct}$. 
This pairwise additive functional form provides a good fit to the target ALMO-EDA data.

The atomic non-bonded parameters are predicted by the GNN model through several distinct approaches.
Similar to our previous work on ByteFF, atomic charges $q$ are predicted using a Bond Charge Correction (BCC) model to preserve charge conservation.
Furthermore, a subset of parameters are obtained using a physically-inspired Tkatchenko-Scheffler (TS) scaling model \cite{tkatchenko_accurate_2009}:
\begin{align}
C_{6,i}=C_{6,i}^{\mathrm{free}}\left(\frac{V_{i,\mathrm{eff}}}{V_{i,\mathrm{free}}}\right)^2,\ \alpha_{i}=\alpha_{i}^{\mathrm{free}}\left(\frac{V_{i,\mathrm{eff}}}{V_{i,\mathrm{free}}}\right),\ r_{i}^*=r_{i}^{*,\mathrm{free}}\left(\frac{V_{i,\mathrm{eff}}}{V_{i,\mathrm{free}}}\right)^{4/21}.
\end{align}
In this model, $V_{i,\mathrm{eff}}$ is the effective atomic volume, predicted from the GNN's atomic hidden states via a Multi-Layer Perceptron (MLP) to account for the local chemical environment.
The terms $C_{6,i}^{\mathrm{free}}$, $\alpha_{i}^{\mathrm{free}}$, $r_{i}^{*,\mathrm{free}}$, and $V_{i,\mathrm{free}}$ are parameters for the free atom, dependent only on its element type.
Finally, the remaining non-bonded parameters ($\epsilon^\mathrm{rep}$, $\lambda^\mathrm{rep}$, $\epsilon^\mathrm{ct}$, and $\lambda^\mathrm{ct}$) are predicted directly by MLPs that take the atom hidden states as input.
We found that incorporating this physically-grounded TS model enhances training stability and model transferability without a noticeable loss of accuracy.

Once all atom-wise parameters are determined, the pairwise interaction parameters are derived using combination rules.
We employ the standard Lorentz-Berthelot (LB) rules \cite{lorentz_ueber_1881}, which use an arithmetic mean for $r_i^*$ and a geometric mean for all other parameters:
\begin{align}
r_{ij}^*=\frac{r_i^*+r_j^*}{2}, \quad \epsilon_{ij}=\sqrt{\epsilon_{i}\epsilon_{j}}, \quad \lambda_{ij}=\sqrt{\lambda_{i}\lambda_{j}}, \quad C_{6,ij}=\sqrt{C_{6,i} C_{6,j}}.
\end{align}

\subsection{Dataset Curation}

ByteFF-Pol's training relies on three core datasets: the existing optimization and torsion datasets from our previous work and a newly created dimer dataset.
Details on the original optimization and torsion datasets are in our previous publication \cite{zheng2025data}.
For this work, we've augmented the optimization dataset by calculating the Minimal Basis Iterative Stockholder (MBIS)~\cite{verstraelen2016minimal} charges and volumes for each molecule at the PBE0/def2-TZVPD level, which are used to pre-train the model.

To capture intermolecular interactions, we curated a large-scale dataset of molecular dimers with ALMO-EDA labels.
This dataset contains 60,790 unique molecular dimers, covering nine element types that are prevalent in organic molecules and electrolytes (C, H, O, N, P, S, F, Cl and Li).
For each dimer, 100 conformations were sampled, resulting in approximately 6 million conformations in total.
The specific methods for constructing these molecular dimers and their conformations are described in Supplementary Note 4.
We performed the second-generation ALMO-EDA calculations at the $\omega$B97M-V/def2-TZVPD level to generate the decomposed interaction energy labels for each conformation.
All the QM calculations were performed using Q-Chem 6.1~\cite{Epifanovsky2021}.
We have also implemented the second-generation ALMO-EDA with GPU4PySCF~\cite{doi:10.1021/acs.jpca.4c05876,wu2025enhancing} for validation (see Supplementary Note 5 for implementation details).

To benchmark the performance of ByteFF-Pol, we used five datasets of experimental physical properties.
The first contains 96 pure molecular liquids with known densities and evaporation enthalpies from the QUBEKit benchmark \cite{hortonQUBEKitAutomatingDerivation2019}.
The second and third datasets consist of experimental densities for 595 binary liquids from the NIST ThermoML archive~\cite{chirico2003thermoml} and 2132 pure molecular liquids from the CRC Handbook~\cite{Lide2004CRC}, respectively.
For electrolyte-relevant systems, we compiled two additional datasets collected from literature~\cite{yang2025unified}: one with 88 experimental densities and 91 experimental viscosities for pure solvents and lithium electrolytes, and another containing 4896 conductivity measurements for various lithium electrolytes across different compositions and temperatures.
The number of results for each force field may differ from that of the benchmark set, owing to parametrization errors or MD simulation failures.

\subsection{Training Techniques}

Our training procedure consists of three stages: pre-training, main training, and fine-tuning.
Detailed training techniques are provided in Supplementary Note 7; here, we present a brief overview.

In the initial pre-training stage, the GNN model is trained to reproduce the bonded parameters from ByteFF~\cite{zheng2025data}, along with atomic charges and volumes from MBIS calculations.
This stage allows the model to develop atomic hidden states that recognize the chemical environment and provides a physically-grounded starting point for the subsequent training.

In the main training stage, the 6 million data points are split into 5.4 million for training and 0.6 million for validation.
We freeze most of the model and train only the output layers corresponding to the non-bonded interaction parameters.
The training targets are the total interaction energies and their decomposed components from the ALMO-EDA calculations.
We employ a scaled mean squared error (SMSE) loss for training each energy term:
\begin{align}
    \mathcal{L}_{\mathrm{SMSE}}^{\mathrm{term}} = \frac{1}{N} \sum_{i} s_i^\mathrm{Boltz} s_i^\mathrm{Force} \left( U^\mathrm{FF, inter}_{\mathrm{term},i} - U^\mathrm{EDA}_{\mathrm{term},i} \right)^2,
    \label{eq:SMSE}
\end{align}
where the symbol term refers to a specific interaction term, i.e., dispersion, electrostatics with Pauli repulsion, polarization and charge transfer.
The two scaling factors, $s_i^\mathrm{Boltz}$ and $s_i^\mathrm{Force}$, reduce the weighting of data points with large interaction energies and large interaction forces, respectively. 
This ensures a stable training process and improves the accuracy for conformations that have higher Boltzmann sampling probabilities.
Detailed definitions of the scaling factors and the decomposed energy terms are provided in Supplementary Note 7.2.

Finally, in the fine-tuning stage, we update the torsion parameters to ensure they are consistent with the newly trained non-bonded interactions.
The training target for this stage is the set of torsion scan energy profiles from ByteFF torsion dataset~\cite{zheng2025data}.

\subsection{Molecular Dynamics Simulation Details}

The macroscopic properties including density ($\rho$), evaporation enthalpy ($\Delta H_{\mathrm{vap}}$), viscosity ($\eta$), conductivity ($\sigma$), static dielectric constant ($\varepsilon(0)$), volumetric expansion coefficient ($\alpha_p$) and isothermal compressibility ($\kappa_\mathrm T$) were computed from the MD simulations.
These simulations were executed using the OpenMM software package~\cite{eastman2017openmm} on L20 GPU cards.
A modified version of OpenMM was utilized to implement the functional forms described in this work (see Code Availability).
Different force fields, including OPLS-AA, AMOEBA and ByteFF-Pol were simulated under the same protocols.
To enhance computational efficiency, the r-RESPA multiple time step Langevin integrator~\cite{zhou1995new} was employed for most simulations, with a time step of 2 fs for non-bonded interactions and 1 fs for bonded interactions, unless otherwise specified.
For the evaluation of density and evaporation enthalpy, we constructed simulation systems containing approximately 5,000 atoms, corresponding to the experimental components; for other properties, simulation systems with around 10,000 atoms were employed.
Detailed simulation protocols and discussions are provided in Supplementary Note 8 and 12.

\section{Data availability}
Trained model checkpoints and validation data shown in Figure 2(a) have been deposited in Hugging Face under accession code \url{https://huggingface.co/ByteDance-Seed/byteff2}, under the CC BY-NC 4.0 license.
A minimal representative training dataset along with atomic coordinates and QM calculation results shown in Figure 2 and Supplementary Figures 1--11 are provided in GitHub under accession code \url{https://github.com/ByteDance-Seed/byteff2/tree/v1.0.0}.
The source data underlying Figures 2--4 and Supplementary Figures 11--21 are provided as Source Data files.
Source data are provided with this paper.

\section{Code availability}
The source code for ByteFF-Pol, including scripts for software installation, training data generation, and the core training scheme, is publicly available in the following GitHub repository: \url{https://github.com/ByteDance-Seed/byteff2/tree/v1.0.0}. This repository also contains comprehensive examples for validation and simulation. The code is provided under the Apache-2.0 license.

\section*{Acknowledgments}
We thank Dr. J. Harry Moore and the other authors of MACE-OFF for providing the detailed simulation results. We thank Dr. Xiaoyu Wang and the other authors of MACELES-OFF for providing the detailed simulation results.

\section*{Author Contributions Statement}
Conceptualization: T.Z., Z.W. and W.Y.; data curation: T.Z. and X.H.; methodology: T.Z., X.X., Z.W., Z.Y. and K.Y.; software: T.Z., X.X., Z.W., Z.Y., Y.W., X.H., L.C., Z.M. and Z.Z.; discussion: Z.W., Z.Y., Z.M., S.L., S.G. and K.Y.; writing: T.Z., X.X., Z.W., X.H., Y.W. and K.Y; supervision: K.Y. and W.Y.

\section*{Competing Interests Statement}
The authors declare that they have no competing financial interests.

\end{document}